\documentclass[runningheads]{llncs}

\usepackage{listings}
\usepackage{caption, subcaption}
\usepackage{graphicx}
\usepackage{cite}
\PassOptionsToPackage{hyphens}{url}\usepackage{hyperref}

\begin{document}

\title{Cited But Not Archived: Analyzing the Status of Code References in Scholarly Articles}

\titlerunning{Cited But Not Archived}

\author{Emily Escamilla\inst{1}\orcidID{0000-0003-3845-7842} \and
Martin Klein\inst{2}\orcidID{0000-0003-0130-2097} \and
Talya Cooper\inst{3}\orcidID{0000-0003-4241-6330} \and
Vicky Rampin\inst{3}\orcidID{0000-0003-4298-168X} \and 
Michele C. Weigle\inst{1}\orcidID{0000-0002-2787-7166} \and
Michael L. Nelson\inst{1}\orcidID{0000-0003-3749-8116}}

\authorrunning{E. Escamilla et al.}

\institute{Old Dominion University, Norfolk, VA, USA \\
\email{evogt001@odu.edu, \{mweigle, mln\}@cs.odu.edu} \and
Los Alamos National Laboratory, Los Alamos, NM, USA \\
\email{mklein@lanl.gov} \and
New York University, New York, NY, USA \\
\email{\{tc3602, vicky.rampin\}@nyu.edu}
}

\maketitle

\begin{abstract}
One in five arXiv articles published in 2021 contained a URI to a Git Hosting Platform (GHP), which demonstrates the growing prevalence of GHP URIs in scholarly publications. However, GHP URIs are vulnerable to the same reference rot that plagues the Web at large. The disappearance of software hosting platforms, like Gitorious and Google Code, and the source code they contain threatens research reproducibility. Archiving the source code and development history available in GHPs enables the long-term reproducibility of research. Software Heritage and Web archives contain archives of GHP URI resources. However, are the GHP URIs referenced by scholarly publications contained within the Software Heritage and Web archive collections? We analyzed a dataset of GHP URIs extracted from scholarly publications to determine (1) is the URI still publicly available on the live Web?, (2) has the URI been archived by Software Heritage?, and (3) has the URI been archived by Web archives? Of all GHP URIs, we found that 93.98\% were still publicly available on the live Web, 68.39\% had been archived by Software Heritage, and 81.43\% had been archived by Web archives.

\keywords{Web Archiving; Digital Preservation; Open Source Software; Memento; Software Heritage}

\end{abstract}

\section{Introduction}

A growing number of researchers reference Git Hosting Platforms (GHPs), like GitHub, GitLab, SourceForge, and Bitbucket, in scholarly publications \cite{escamilla-tpdl2022}. GHPs are Web-based hosting platforms for git repositories. GHPs are commonly used by software developers, including researchers, to host software and facilitate collaboration. Researchers include GHP URIs in their publications for software products that were either used in their research or created in the course of the study. However, scholarly code products hosted on the live Web with GHPs are susceptible to the reference rot that plagues the Web as a whole \cite{klein-plos2014}. When navigating to a URI found in a publication, users may find a ``404: Page not found'' error due to either the hosting platform or the software repository being no longer available at the URI. 

GHPs provide access to a repository for the lifespan of the GHP, but the long term access provided by preservation is not a priority for GHPs, as shown by the discontinuation of the GHPs Gitorious \cite{gitorious} and Google Code \cite{google_code}. All URIs to those two platforms in scholarly publications no longer point to the content that the author originally intended, due to no fault of the author. The disappearance of these resources creates a problem for scholars interested in replicating the results as well as those interested in the context of the research findings. A reader may be able to successfully locate one of these now-defunct URIs in a Web archive, like the Internet Archive (IA), however, Web archives do not give priority to archiving software products. Therefore, there is no guarantee that any given URI will be preserved, unless a user submits a repository with IA's `Save Page Now'\footnote{\url{https://web.archive.org/save/}} feature.

Software Heritage is a non-profit organization that works ``to collect, preserve, and share all software that is publicly available in source code form'' \cite{swh-mission}. They utilize a `Save Code Now' feature, as well as automated crawling, to create snapshots of origin URIs \cite{dicosmo-ipres2017, dicosmo-icms2020}. While Web archives typically have a large scope covering a wide variety of content types, Software Heritage is singularly focused on the archival of source code and its development history. Software Heritage announced that 202,254,500 projects have been archived as of January 17, 2023.\footnote{\url{https://twitter.com/SWHeritage/status/1615422314224701440}} But, are the software products that scholars are referencing in their publications included in the over 202 million projects that have been captured? 

In this paper, we identified the current state of scholarly code products on the live Web as well as in archives such as Software Heritage and Web archives. We analyzed GHP URIs from 2.6 million articles in the arXiv and Pub Med Central (PMC) corpora and found that 93.98\% of all GHP URIs referenced were publicly available on the live Web. We also found that 68.39\% of all repository URIs were captured by Software Heritage and 81.43\% of all GHP URIs referenced in scholarly publications had at least one archived version in a Web archive. 

\section{Related Work}
In a previous study \cite{escamilla-tpdl2022}, we studied the prevalence of GHP URIs within scholarly publications. Our corpus contained over 2.6 million publications from arXiv and PMC. We extracted 7.7 million URIs from the articles including 253,590 URIs to four GHPs: GitHub, GitLab, Bitbucket, and SourceForge. We found that, over time, a growing number of papers referenced GHP URIs, with 20\% of 2021 arXiv publications containing a GitHub URI. We also found that 33.7\% of publications that reference a GHP include more than one GHP URI. Scholars include URIs to resources that contributed to or impacted their research. The increasing inclusion of GHP URIs in scholarly publications points to an increased importance of the holdings of GHPs for reproducibility in research.


However, URIs to GHPs, like the Web at large, may experience reference rot. Reference rot refers to the occurrence of content drift and link rot \cite{vandesompel-icm2014}. A reference has experienced content drift when the content of the URI when it was referenced is different than the content shown to the user at present. A reference has experienced link rot when the URI is completely inaccessible on the live Web. Link rot may cause the ``404: Page not found'' error that most users have experienced. In a study on the use of URIs to the Web at large in the arXiv, Elsevier, and PMC corpora, Klein et al. \cite{klein-plos2014} found that reference rot affects 20\% of Science, Technology, and Math (STM) publications. In a study on the same corpus, Jones et al. \cite{jones-plos2016} found that 75\% of URI references experienced content drift, with content drift worsening over time. 

Some scholars take an active role in preserving their software, a strategy known as self-archiving. Self-archiving puts the responsibility on scholars to deposit their code product into a repository that guarantees long-term preservation, like Zenodo \cite{peters_zenodo} or the Open Science Framework \cite{foster_osf}. However, a study by Milliken et al. found that only 47.2\% of the academics who create software products self-archive their software \cite{iasge_enviro_scan}. 

Software Heritage preserves source code and its development history from the perspective that source code is itself a valuable form of knowledge that should be captured, including the unique evolution of the source code to create the code product at a given point in time \cite{dicosmo-ipres2017}. Software Heritage provides a central repository containing the source code and development histories of millions of code products across programming languages, hosting platforms, and package repositories. The result is a repository that researchers can use as a more representative sample than a single hosting platform or package library to analyze source code. For instance, Pietri et al. \cite{pietri-msr2020} and Bhattacharjee et al. \cite{bhattacharjee-msr2020} leverage the scope of the Software Heritage dataset to analyze trends in software development across a more heterogeneous dataset than could be found in a single hosting platform. While studies like these have made use of the holdings of Software Heritage, they do not analyze what has or has not been preserved in Software Heritage.

\section{Methodology}
We used a dataset of GHP URIs from our previous study \cite{escamilla-tpdl2022}, where we extracted GitHub, GitLab, Bitbucket, and SourceForge URIs from a corpora of 2,641,041 arXiv and PMC articles. In total, the dataset contained 253,590 GHPs URIs that were referenced in scholarly publications. The distribution of the URIs in each GHP is shown in Table \ref{tab:ghps}. For each URI in the dataset, we conducted three tests: (1) is it available on the live Web?, (2) is it available in Software Heritage?, (3) is it available in Web archives? We also analyzed the relationship between the publication date of the earliest article to reference a URI and the date of the first Software Heritage and Web archive capture of the URI\footnote{Source code and datasets are available at \url{https://github.com/oduwsdl/Extract-URLs}}.


\begin{table}
    \centering
    \begin{tabular}{|l|r|r|}
    \hline
    GHP & Number of URIs & Percent of GHP URIs\\ 
    \hline
    GitHub & 234,092 & 92.31\% \\
    SourceForge & 12,721 & 5.01\% \\
    Bitbucket & 3,962 & 1.56\% \\
    GitLab & 2,815 & 1.11\% \\
    \hline
    \end{tabular}
    \caption{Number of URIs to each GHP and the percentage of all GHP URIs}
    \label{tab:ghps}
\end{table}

In this study, we adopt the terminology used by Klein et al. \cite{klein-plos2014}. A URI is \emph{publicly available} if a curl request results in a 2XX-level HTTP response code. If a URI is publicly available, we consider the URI to be \textbf{active} on the live Web. Private repositories respond to a curl request with a 404 HTTP response code. While a private repository exists and is available to the owner, it is not publicly available and accessible via the URI provided in the scholarly publication; therefore, because a URI to a private repository is not available to general users, it is not considered an active URI. Any URI that does not result in a 2XX-level HTTP response code is considered inactive, meaning that the URI is \textbf{rotten}, or is subject to link rot. In our curl requests, we opted to follow redirects and considered the resulting HTTP response code as the final status of the URI on the live Web. 

We utilized the Software Heritage API \cite{swh-api} to determine if Software Heritage contained a snapshot of the URI. However, Software Heritage only supports searching for URIs at the repository level, whether through their browser search interface or API request. Searching for deep links to a specific file or directory will not result in a match, even if the file or directory is available within Software Heritage's snapshot of the repository. For example, \url{https://github.com/aliasrobotics/RVD/blob/master/rvd\_tools/database/schema.py} is a URI that was extracted from an article in the arXiv corpus. As it is written, the Software Heritage API was not able to find a matching origin. When we truncate the URI to the repository-level (\url{https://github.com/aliasrobotics/RVD}), the Software Heritage API returned a matching origin URL. To accommodate the requirements of the Software Heritage API, we transformed all deep URIs to GitHub, GitLab, and Bitbucket to shallow, repository level URIs and requested the resulting URI from the Software Heritage API. SourceForge affords a high level of customization for hosted projects. SourceForge projects can support issue pages, documentation, wiki pages, and code repository pages. However, these pages are not required. As a results, not all SourceForge projects have publicly available source code that can be cloned via an access URL. Software Heritage utilizes access URLs to capture and preserve source code. As a result, SourceForge projects that do not provide an access URL are not able to be captured by Software Heritage. In order to accurately reflect the archival rate of SourceForge URIs in Software Heritage, we excluded SourceForge URIs to projects that did not provide an access URL. To determine if the project provided an access URL, we sent a request to the SourceForge API with the project name. From the SourceForge API response, we extracted all access URLs. Of 7,269 unique SourceForge projects, 47.08\%, or 3,422 projects did not provide at least one access URL. If the project did not contain any access URLs, the project URI was excluded. Otherwise, if the project contained one or more access URLs, we requested all access URLs for the SourceForge project from the Software Heritage API. The Software Heritage API returned metadata for each snapshot including the origin (the original URI and the type of software origin), visit number, date of the snapshot, status of the snapshot, and the snapshot ID \cite{dicosmo-ipres2017}. From the API response, we extracted the date of the first and last snapshot and the total number of snapshots for each URI. 

We used MemGator \cite{jcdl-alam-memgator}, a Memento \cite{RFC7089} aggregator that queries the holdings of 12 distinct Web archives, to search for the GHP URIs. The result of a Web archive crawling a live Web page, identified by a URI-R, is a \emph{memento}, an archived version of the URI-R at a given point in time and is identified by a URI-M. After requesting the URI from each of the Web archives, MemGator compiles all of the archives' responses for the URI-R into a TimeMap that includes the URI-M of each memento and the corresponding Memento-Datetime (i.e., the date it was archived). From the resulting TimeMap, we extracted the Memento-Datetime of the first and last memento and the total number of mementos for each GHP URI-R.  

McCown and Nelson \cite{mccown-jcdl2009} developed a framework for discussing the intersection of Web archiving and the life span of a Web resource, which we have adapted to discuss the intersection of Web archiving and the life span of source code in a GHP. We define a GHP URI resource as \textbf{vulnerable} if it is publicly available on the live Web but has not been archived. If a GHP URI resource is publicly available on the live Web and has been archived, we define the GHP URI resource as \textbf{replicated}. Lastly, we define a GHP URI resource as \textbf{unrecoverable} if it is no longer publicly available on the live Web and has not been archived. 

\section{Results}
\begin{figure}
\centering
\begin{subfigure}{0.8\textwidth}
    \includegraphics[width=0.95\linewidth]{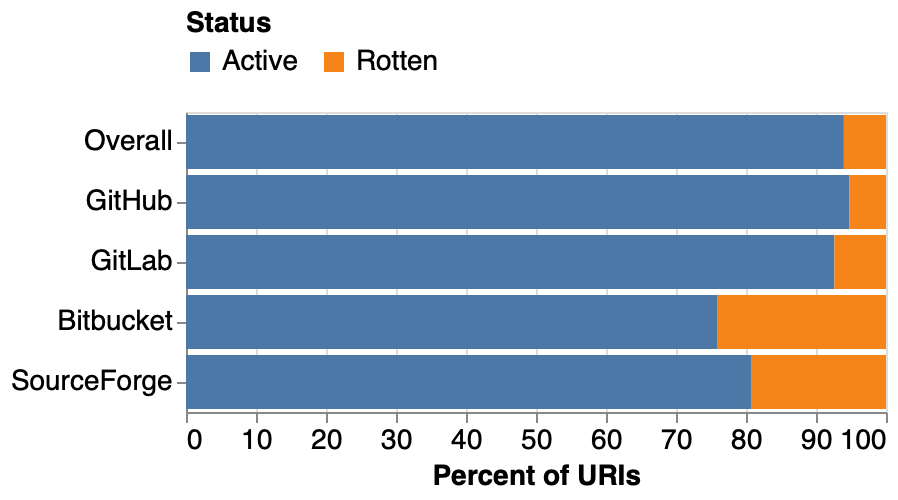}
    \caption{Percent of active URIs}
    \label{fig:alive}
\end{subfigure}
\begin{subfigure}{0.8\textwidth}
    \includegraphics[width=0.95\linewidth]{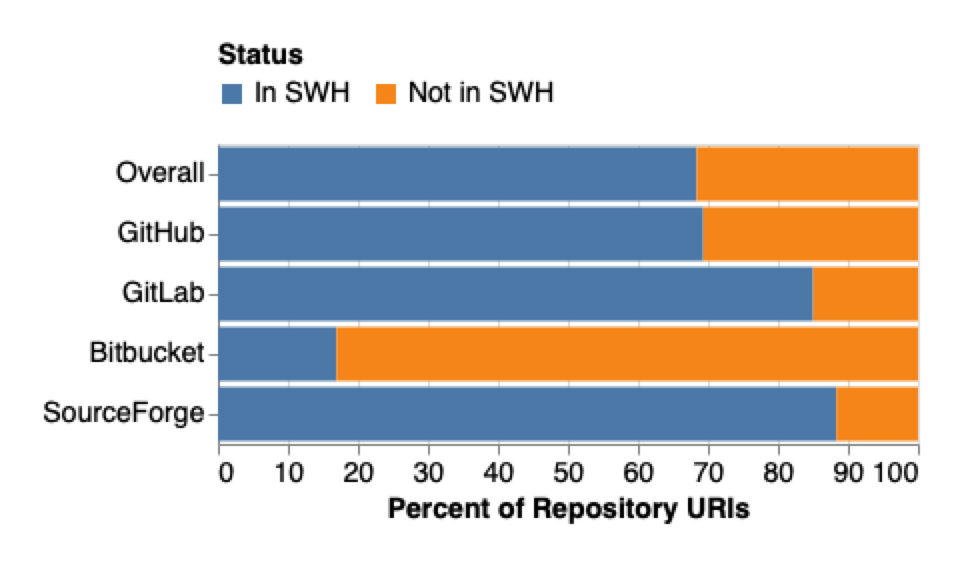}
    \caption{Percent of repository URIs captured by Software Heritage (SWH)}
    \label{fig:swh}
\end{subfigure}
\begin{subfigure}{0.8\textwidth}
    \includegraphics[width=0.95\linewidth]{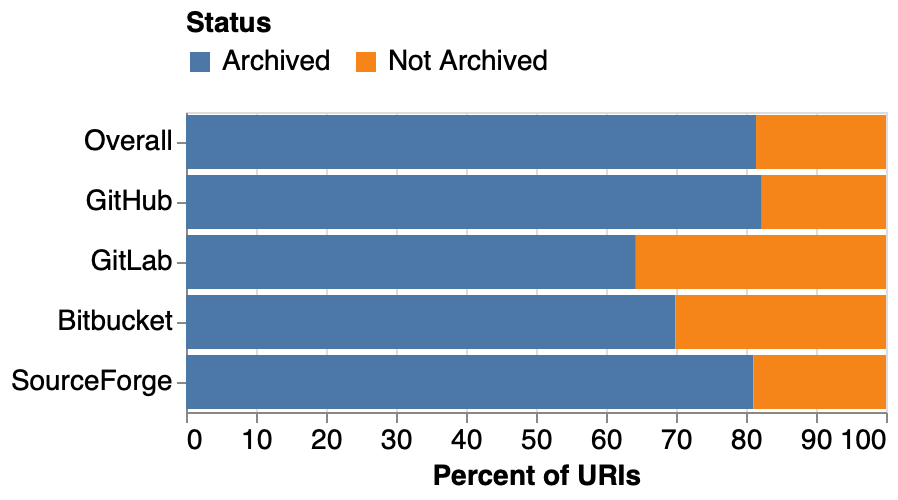}
    \caption{Percent of URIs that had at least one memento}
    \label{fig:timemap}
\end{subfigure}
\caption{Results of running the three tests: (1) is the URI active?, (2) has the URI been archived by Software Heritage?, and (3) has the URI been archived by Web archives?}
\label{fig:is_it}
\end{figure}

Across all four GHPs, 93.98\% of all GHP URIs referenced in scholarly publications were active, as shown in Figure \ref{fig:alive}. However, 6.02\%, or 8,882 URIs of the unique GHP URIs in scholarly publications, were rotten. GitHub had the highest percentage of active URIs with 94.79\%. Bitbucket had the lowest percentage of active URIs with 75.86\% resulting in 641 rotten URIs. 

As shown in Figure \ref{fig:swh}, 68.39\% of all repository-level GHP URIs have at least one snapshot in Software Heritage. SourceForge had the highest percentage of repository URIs captured by Software Heritage with 88.34\%. Bitbucket had the lowest percentage with 16.93\% of repository URIs.

Across all four GHPs, 81.43\% of GHP URIs have at least one memento in the Web archives queried by MemGator, as shown in Figure \ref{fig:timemap}. GitHub had the highest percentage of URIs available in Web archives with 82.25\% and SourceForge was a close second with 81.06\%. GitLab had the smallest percentages of URIs available in Web archives with 64.29\%. The distribution of the percent of mementos returned from each of the twelve Web archives is shown in Table \ref{tab:archives}. Internet Archive had the largest percent of all returned mementos with 58.68\%. Internet Archive is followed by Bibliotheca Alexandrina Web Archive,\footnote{\url{https://www.bibalex.org/isis/frontend/archive/archive\_web.aspx}} which returned 23.29\% of all mementos. However, we note that since 2022 Bibliotheca Alexandrina has functioned as a backup to the Internet Archive and provides a mirror of the Internet Archive's holdings \cite{bibalex-ia}. This could explain the high percentage of GHP URIs available in both the Internet Archive and Bibliotheca Alexandrina Web archives. The remaining 18.03\% of mementos are distributed across the remaining 10 Web archives. 

\begin{table}
    \centering
    \begin{tabular}{|l|c|}
    \hline
    Web Archive & Percent of Mementos \\
    \hline
    Internet Archive & 58.68\% \\
    Bibliotheca Alexandria Web Archive & 23.29\% \\
    Archive.today & 8.06\% \\
    Archive.it & 3.07\% \\
    Portuguese Web Archive & 2.83\% \\
    Library of Congress & 2.53\% \\
    Icelandic Web Archive & 0.88\% \\
    Australian Web Archive & 0.35\% \\
    UK Web Archive & 0.12\% \\
    Perma & 0.11\% \\
    Stanford Web Archive & 0.08\% \\
    BAnQ & 0.0005\% (1 URI-M) \\ \hline
    \end{tabular}
    \caption{Percent of all mementos returned from each of the 12 Web archives}
    \label{tab:archives}
\end{table}

\begin{figure}
\centering
\begin{subfigure}{0.9\textwidth}
    \includegraphics[width=0.95\linewidth]{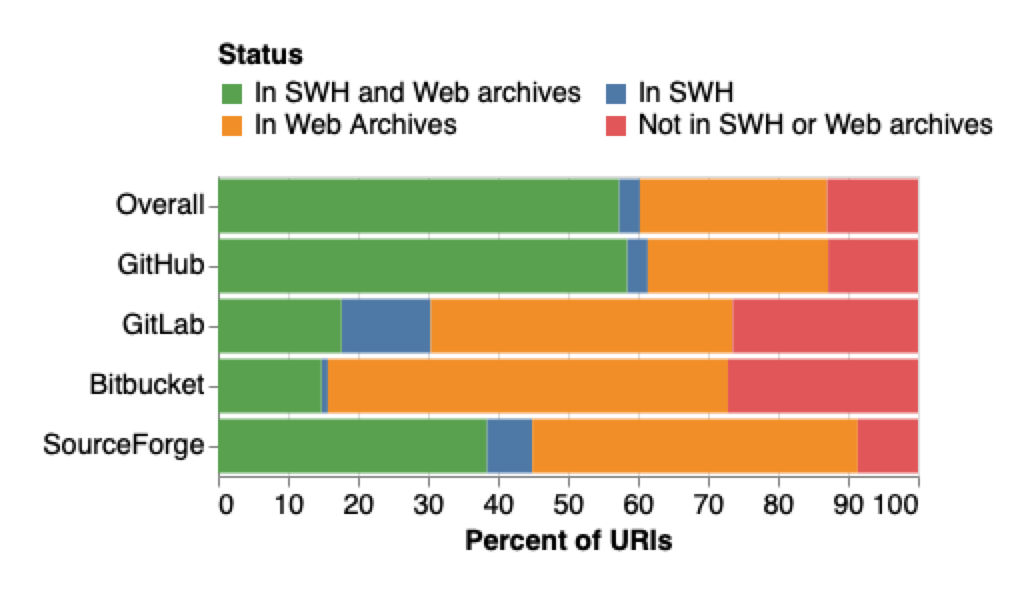}
    \caption{Percent of repository-level URIs overall and for each GHP}
    \label{fig:who_has_a_copy}
\end{subfigure}
\begin{subfigure}{0.9\textwidth}
    \centering
    \includegraphics[width=0.95\linewidth]{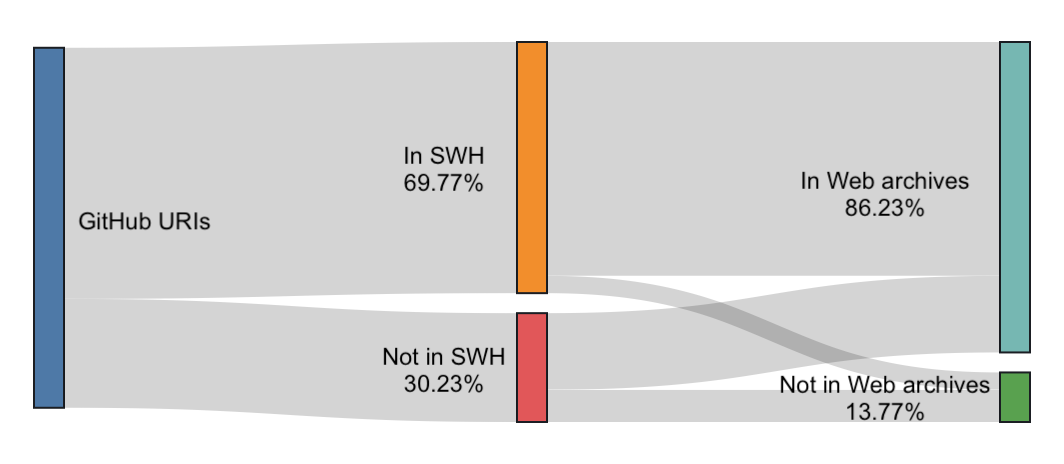}
    \caption{Relationship between the number of GitHub URIs preserved}
    \label{fig:gh_sankey}
\end{subfigure}
\begin{subfigure}{0.9\textwidth}
    \centering
    \includegraphics[width=0.95\linewidth]{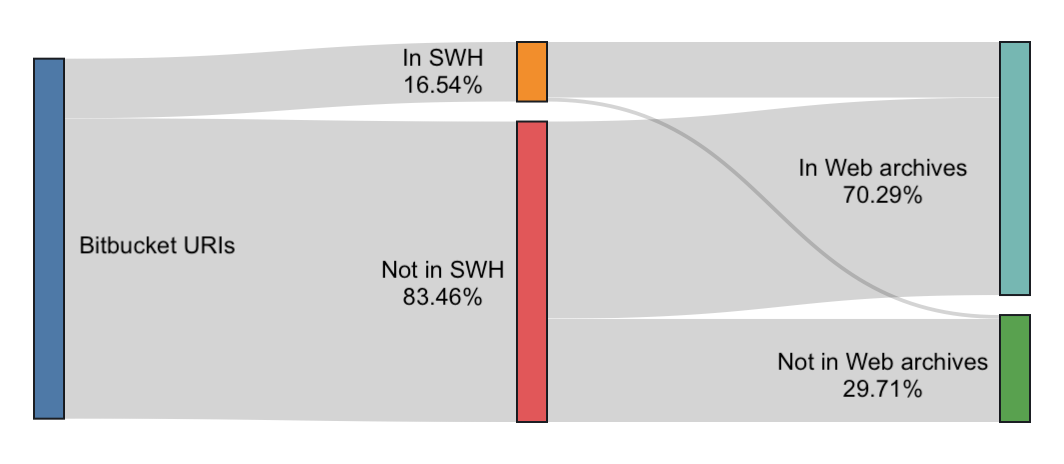}
    \caption{Relationship between the number of Bitbucket URIs preserved}
    \label{fig:bb_sankey}
\end{subfigure}
\caption{Relationships between URIs that have been archived by Software Heritage (SWH) and Web archives, only Software Heritage, only Web archives, and neither Software Heritage or Web archives}
\end{figure}

Figure \ref{fig:who_has_a_copy} depicts the percent of URIs archived by both Software Heritage and Web archives, only Software Heritage, only Web archives, and neither Software Heritage or Web archives. Overall, 57.21\% (30,311 URIs) of all URIs were captured by both Software Heritage and Web archives and 12.99\% (5,249 URIs) were not captured by either Software Heritage or the Web archives, making their resources vulnerable. Across all four GHPs, there are a higher percentage of GHP URIs that have only been archived by Web archives (26.74\%) than the percentage of GHP URIs that have only been archived by Software Heritage (3.02\%). SourceForge and Bitbucket have the highest percentage of URIs unique to the Web archives with 46.50\% of SourceForge URIs and 57.15\% of Bitbucket URIs only archived by Web archives. Bitbucket has the highest percentage of vulnerable URI resources with 27.15\%. Figures \ref{fig:gh_sankey} and \ref{fig:bb_sankey} give a more detailed look at the relationship between each category for GitHub and Bitbucket URIs.  

As shown in Figure \ref{fig:gh_sankey}, 58.43\% of GitHub URIs have been archived by both Software Heritage and Web archives, while 25.76\% of GitHub URIs have only been archived by Web archives. These percentages noticeably differ from the distribution of Bitbucket URIs as depicted in Figure 
\ref{fig:bb_sankey}. Of all Bitbucket URIs, 14.78\% have been archived by both Software Heritage and Web archives while 57.15\% have only been archived by Web archives. Additionally, only 0.91\% of Bitbucket URIs are archived by Software Heritage and not archived by Web archives, compared to 2.97\% of GitHub URIs. 

\begin{figure}
\centering
\begin{subfigure}{0.8\textwidth}
    \centering
    \includegraphics[width=\linewidth]{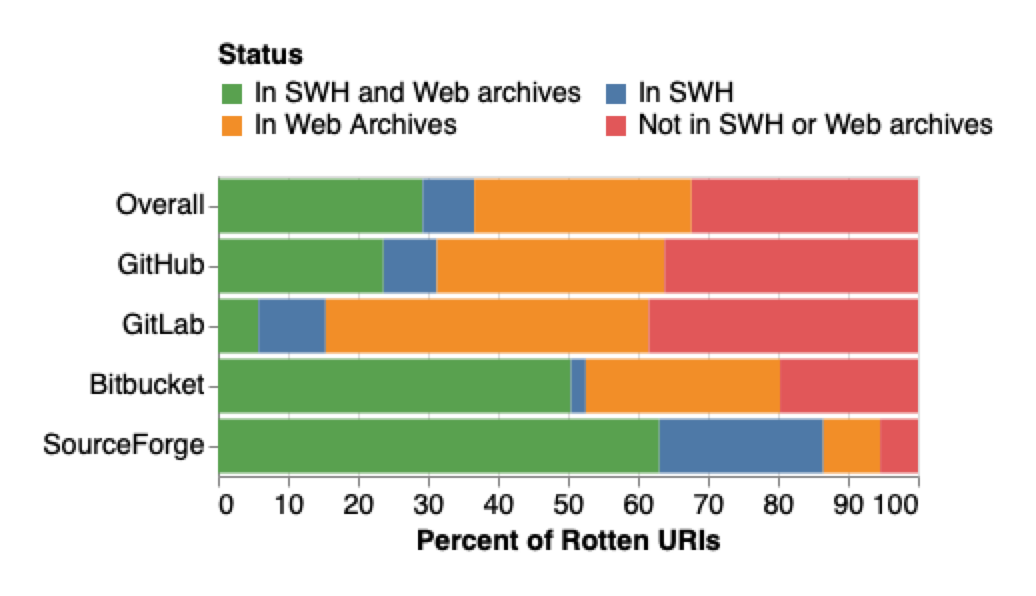}
    \caption{Percent of rotten repository-level URIs overall and for each GHP}
    \label{fig:dead_and_gone}
\end{subfigure}
\centering
\begin{subfigure}{0.8\textwidth}
    \centering
    \includegraphics[width=\linewidth]{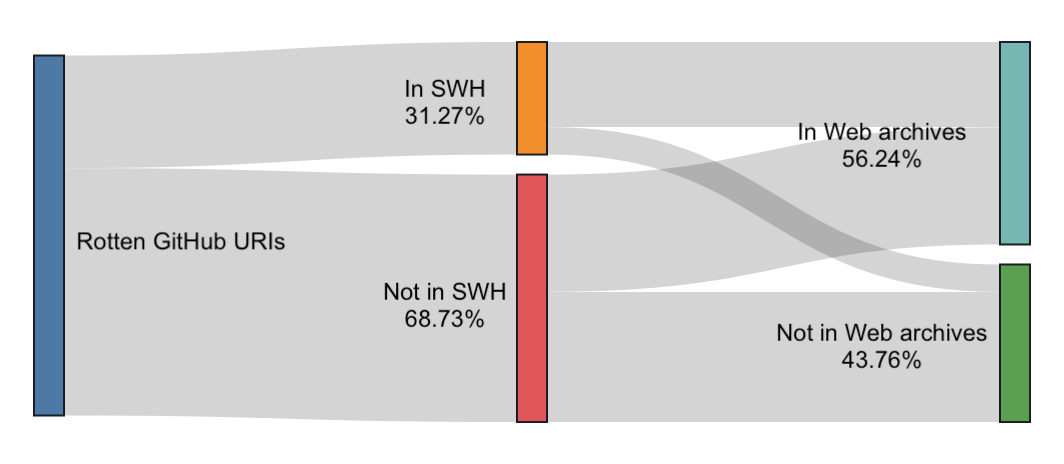}
    \caption{Preservation of rotten GitHub URIs}
    \label{fig:gh_dead_sankey}
\end{subfigure}
\begin{subfigure}{0.8\textwidth}
    \centering
    \includegraphics[width=\linewidth]{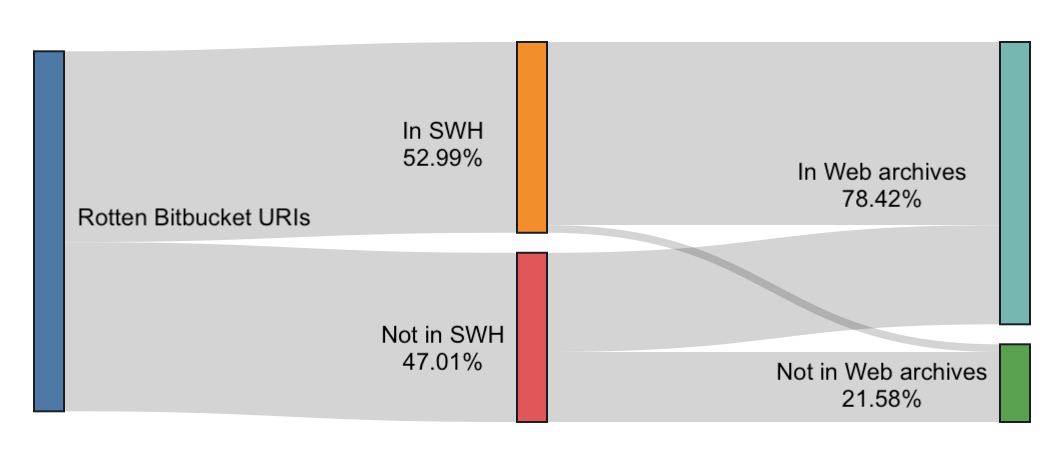}
    \caption{Preservation of rotten Bitbucket URIs}
    \label{fig:bb_dead_sankey}
\end{subfigure}
\caption{Relationships between rotten URIs that have been archived by Software Heritage (SWH) and Web archives, only Software Heritage, only Web archives, and neither Software Heritage or Web archives}
\end{figure}

Because they have active URIs, vulnerable resources still have the opportunity to be archived by Web archives and Software Heritage. However, rotten URIs are no longer able to be preserved. Figure \ref{fig:dead_and_gone} depicts the percent of rotten URIs that have been archived by both Software Heritage and Web archives, only Software Heritage, only Web archives, and neither Software Heritage or Web archives. Across all four GHPs, 2,823 unique repository-level URIs are rotten and 32.36\% of those rotten URIs are unrecoverable, as they have not been archived by Software Heritage or Web archives. In total, there are 932 unrecoverable URI resources across all four GHPs. GitLab has the smallest percentage of rotten URIs that have been archived by both Software Heritage and Web archives with 5.77\%. Conversely, SourceForge has the largest percentage of rotten URIs that are captured by both Software Heritage and Web archives with 63.06\%. For rotten URIs, there is a smaller percentage of URIs that have only been archived by Software Heritage (7.37\%) than the percentage of URIs that have only been archived by Web archives (31.04\%). This trend is similar to what we saw for all GHP URIs in Figure \ref{fig:who_has_a_copy}. Figures \ref{fig:gh_dead_sankey} and \ref{fig:bb_dead_sankey} provide a more detailed look at the relationship between each category for rotten GitHub and Bitbucket URIs. 

As shown in Figure \ref{fig:gh_dead_sankey}, 36.22\% of rotten GitHub URIs are unrecoverable. Inversely, 23.64\% of rotten GitHub URIs have been archived by both Software Heritage and Web archives. GitHub has a larger percentage of rotten URIs that have only been archived by Software Heritage (7.63\%) than Bitbucket (2.14\%) as shown in Figure \ref{fig:bb_dead_sankey}. Again, the distribution of rotten Bitbucket URIs is distinguishable from the distribution of rotten GitHub URIs. We found that 19.70\% of rotten Bitbucket URIs are unrecoverable while 50.42\% of rotten Bitbucket URIs have been archived by both Software Heritage and Web archives.  

For both Software Heritage and Web archives, we calculated the time between the date of the first publication to reference a URI and the date of the first capture of the URI. Software Heritage was created on June 30, 2016 \cite{dicosmo_swhblog}, so we only analyzed articles that were published starting July 1, 2016. We found an average of 443 days (median of 360 days) between the first reference to the repository URI in a scholarly publication and the first capture by Software Heritage, if the repository-level URI did not have a snapshot at the time of publication. Additionally, 7,440 repository URIs that were captured before the publication date of the referencing article had not been captured since the article's publication. For these  URIs, there is an average of 253 days between the last Software Heritage snapshot and the publication date of the reference article. 

\begin{figure}
    \centering
    \includegraphics[width=0.85\linewidth]{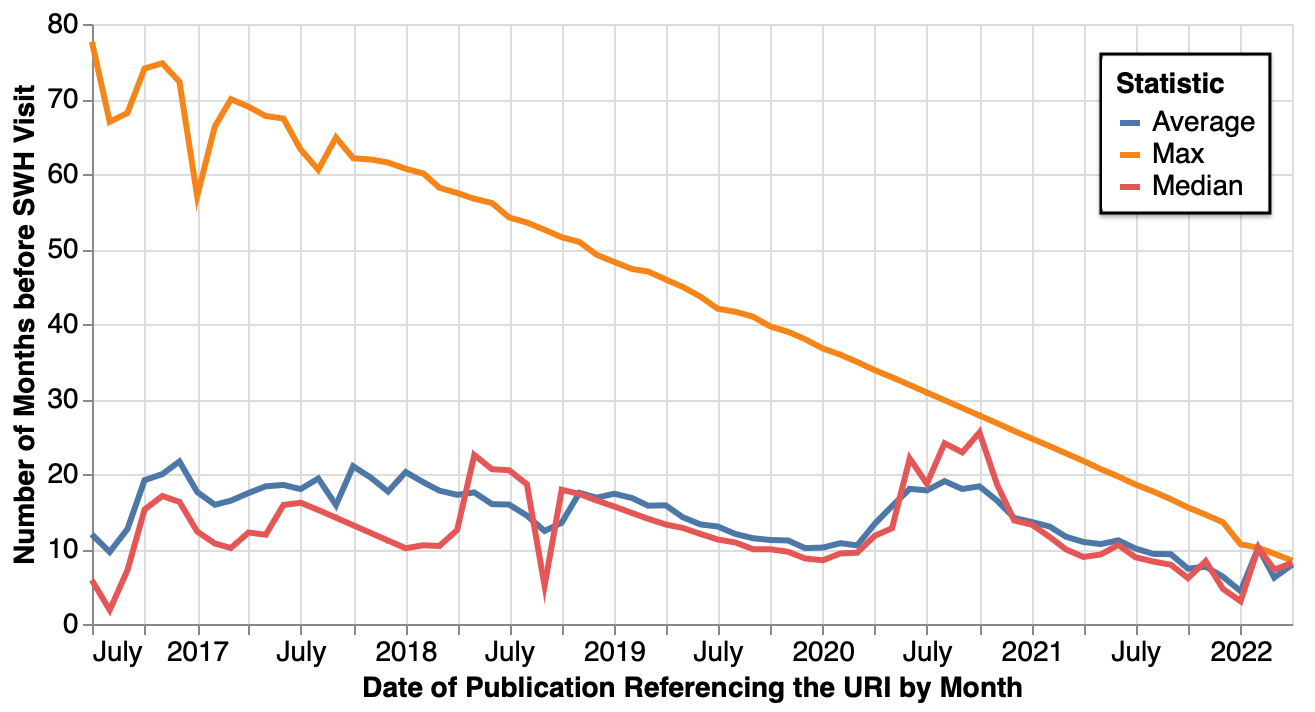}
    \caption{Months between a publication referencing a URI and the URI being captured by Software Heritage over time. Only includes URIs not been captured by Software Heritage before the publication date of the referencing article.}
    \label{fig:swh_delta}
\end{figure}

As shown in Figure \ref{fig:swh_delta},
the maximum time delta between the first reference to the repository URI in a scholarly publication and the first capture by Software Heritage has steadily decreased from 78 months for articles published in July 2016 to 9 months for articles published in April 2022. We also see that the median time delta follows a trend similar to the average time delta. The median and average time deltas have both decreased since 2021. 

\begin{figure}
    \centering
    \includegraphics[width=0.85\linewidth]{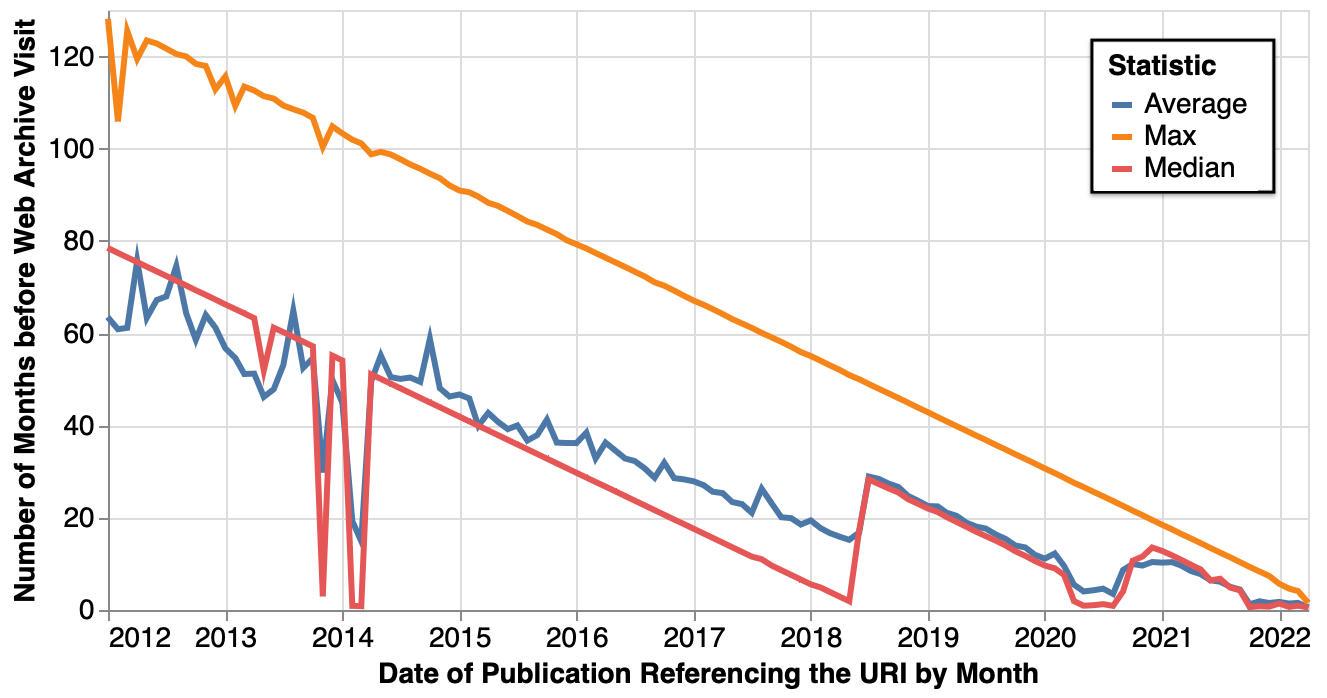}
    \caption{Number of months between a publication referencing a URI and the URI being captured by the Web archives over time. Only includes URIs not captured by the Web archives before the publication date of the referencing article.}
    \label{fig:timemap_delta}
\end{figure}

These trends for are similar for Web archives. There was an average of 468 days and a median of 341 days between the first reference to the URI in a scholarly publication and the first memento in a Web archive, if there were no mementos of the URI prior to the publication date of the referencing article. Of the URIs that had a memento in the Web archives prior to the publication date of the article, 4,356 URIs have not been archived since the article was published, with an average of 201 days between the latest memento and the publication date. Figure \ref{fig:timemap_delta} shows that the average and maximum time deltas have followed similar trends. Additionally, the maximum time delta has steadily decreased from 128 months in January 2012 to 1 month in April 2022. While the steady decline seen in maximum and average time deltas for Software Heritage and Web archives is promising, there is still a large period of time for the URI resource to move from vulnerable to unrecoverable before Software Heritage or Web archives are able to archive it.

\section{Discussion}

We analyzed the GHP URIs that were extracted in a previous study from an arXiv and PMC corpora. GHP URIs from other corpora may produce a different result. For example, authors must proactively submit their paper to arXiv, which demonstrates an inclination to participate in open research. As such, authors who submit to arXiv may be more likely to submit source code projects to Software Heritage and Web archives for preservation and research reproducibility.

The smaller percentage of Bitbucket and SourceForge URIs publicly available on the live Web and preserved in Software Heritage may be correlated to the usage trends we observed in our previous study \cite{escamilla-tpdl2022}. SourceForge was created in 1999, so older publications are more likely to contain a link to SourceForge than to other GHPs. Additionally, Bitbucket was referenced in scholarly publications more than GitHub from 2008 to 2014, which could also result in older publications containing a link to Bitbucket over other GHPs. As Klein et al. found, the likelihood of reference rot grows as the age of the URI increases\cite{klein-plos2014}, which could be reflected in the lower percentage of SourceForge and Bitbucket URIs still publicly available on the live Web. Additionally, older GHP URIs may be less likely to be preserved in Software Heritage given that Software Heritage was launched June 30, 2016. Some of the GHP URIs that are not publicly available on the live Web may have disappeared long before Software Heritage even existed to preserve them. 

For the purposes of this study, we looked at the presence of an archived copy of the repository in Software Heritage and the Web archives, but did not investigate the quality of the copy. In some instances, the memento or capture may reflect that the URI is no longer available which would not be beneficial for reproducibility. In other cases, the capture may be incomplete in a way that negatively impacts reproducibility \cite{brunelle-ijdl2015}. Determining that a archived copy of the repository exists is the first step in utilizing archived repositories to support reproducibility. In future work, we will investigate the quality of the mementos and captures that are currently available.

As we discussed while introducing the current software archival initiatives, the primary goal of Software Heritage is the preservation of software while Web archives work to preserve the Web at large, including a wide variety of content types. Therefore, it was interesting to find that the Web archives have archived almost 24\% more scholarly GHP URIs than Software Heritage. While the repositories captured by Software Heritage are not a perfect subset of the repositories captured by Web archives, the holdings of the Web archives have a more complete coverage of scholarly repositories. A higher level of archival coverage benefits reproducibility as more scholarly software is made available to support long-term reproducibility. 

With 93.98\% of URIs still publicly available on the live Web, there is the opportunity to submit these URIs to be preserved. The 1.98\% of URIs that are rotten and unrecoverable should serve as a warning of what could happen if the research community does not act to preserve code products as integral research products. Researchers need to take initiative to submit code products to services like Software Heritage and Web archives to ensure the code they reference is preserved for long term access. 

\section{Conclusions}
The inclusion of a URI to a GHP in a scholarly publication indicates the importance and impact of the repository to a scholar's research. Research reproducibility hinges on the ability of researchers to access the data and source code that contributed to a research outcome. We found that current archival efforts by Web archives and Software Heritage do not adequately preserve the GHP URIs referenced in scholarly publications. Additionally, Software Heritage, an archive solely focused on the preservation of software, contained fewer scholarly software products than the Web archives. Overall, 68.39\% of the repository URIs were captured by Software Heritage while 81.43\% had at least one memento in the Web archives. We also found 12.99\% of the GHP URIs were not archived in Software Heritage or Web archives and 32.36\% of the GHP URIs that are no longer available on the live Web and not archived in either Software Heritage or the Web archives. 

\bibliographystyle{splncs04}
\bibliography{bib-file}

\end{document}